\documentclass[12pt]{iopart}
\usepackage{epsfig}

\begin{document}
\title[A Teachers' primer to the Introduction of the Cross-section Concept]{Reflection off Surfaces of Revolution:\\
a Teachers' primer to the Introduction of the Cross-section Concept}
\author{Marco Giliberti and Luca Perotti${}^{a}$}
%\affiliation
\address{Sezione Didattica della Fisica, Dipartimento di Fisica dell'Universit\'a di Milano, via Celoria 16, 20133 Milano Italy}
%\affiliation
\address{${}^{a}$Department of Physics, Texas Southern University, Houston, Texas 77004 USA}\date{\today}

\begin{abstract}
The differential cross-section for the reflection
of light beams off rigid bodies obtained by the rotation of a
generic derivable convex function is calculated. The calculation
is developed using elementary notions of calculus and is therefore
suitable for calculus oriented introductory undergraduate university physics courses. Three particular cases are presented
as examples of the general procedure and of the physical properties
and considerations about cross-sections they allow to discuss in class.
\end{abstract}

\maketitle

\section{Introduction}

Cross-section is a fundamental conceptual tool in modern physics.  
As the ``effective surface" the target presents to the probing
radiation, be it material or electromagnetic, the cross-section
contains all the information about the nature of the interactions
between probes and target we can extract from the experiment, and
which can be used to derive information on the target structure in the inverse scattering problem.

Unfortunately the first taste of this concept usually comes to the students in the case of the Geiger-Marsden experiment on the
Rutherford scattering. The calculation of cross-sections for this and other
physically relevant cases, such as deep inelastic scattering in
Nuclear Physics, is usually rather long and often too
difficult for modern undergraduate students; moreover the
Rutherford scattering is a very special case, in that, as the
Coulomb potential is a long-range one, its total cross section is
infinite. Examples useful to clarify the geometric meaning of
cross-section, if given, are therefore generally very few:
typically scattering off rigid spheres, which has the drawback of
being a very special case of isotropic scattering.

To overcome this difficulty, we have experimented with an
approach where the concept of cross section is introduced through a
class of interactions for which it is especially easy to grasp:
that of reflection (according to the rules of geometrical optics)
of light beams off surfaces having axial
symmetry. Although the cross-sections thus calculated are of
limited physical significance (but so are most classical mechanics
examples), the interaction is of a kind closer to the student's
everyday experience than those for physically relevant cases and
can also easily be practically demonstrated in class.

Based on simple geometrical considerations, this approach allows an
explicit calculation of several cross-sections with very little
effort: namely that of finding the inverse function of a derivative
(a similar approach, but within the frame of particle mechanics,
and using a different procedure, can be found in Ref. \cite{brun}).
The application of the formula allows the teacher ample material
for classroom discussions about the concept of cross-section, thus
providing a clear introduction to the concept, useful for further
developments. The mathematics required is standard subject in most
science university courses, but, as we observed
in our in-class experimentation, the introductory physics students are
still ``green" enough in physics that intuitive geometrical
examples as the ones here proposed can be both interesting and
useful in fixing in their minds both the basic concepts and the
difficulties, such as for example the fact that the inverse
scattering problem is an arduous one, and often also an ill-posed
one \footnote{A simple example for the students is given by the Rutherford scattering itself: repulsive and attractive Coulomb potentials give the same cross section.}.
%\cite{coulomb}.

While our original experimentation was conducted within the frame
of particle scattering off rigid bodies, the formulation we present
here is in terms of ``optical" scattering, more suitable to contemporary views in fundamental physics education \cite{caz,gil}. As the arguments used are strictly
geometrical, the difference between the two formulations is just in
the language used and any teacher wishing to present the subject
matter in a traditional course, will have no problem in reverting
to a particle scattering description.

The paper is organized as follows: section \ref{due} introduces
through an example the concept of  total cross-section; section
\ref{tre} introduces the class of interactions we use as a model
and through it illustrates the concept of differential
cross-section; section \ref{qua} outlines the proposed approach to
the calculation of the differential cross-section for the chosen
class of interactions; section \ref{cin} presents three particular
significant cases, namely that of ellipsoids, which reduces in the
case of equal semi-axes to the classical example of spherical
targets; the case of paraboloids, which give the same angular
dependence of the differential cross-section as the Rutherford
experiment and, finally, the case of targets generated by the
rotation of an inverse sine curve which presents a curious
similarity to our second example. In section \ref{sei} we discuss
our approach and summarize the advantages of the proposed approach.

\section{The concept of total cross-section}\label{due}

\begin{figure}[htbp]
\centering\epsfig{file=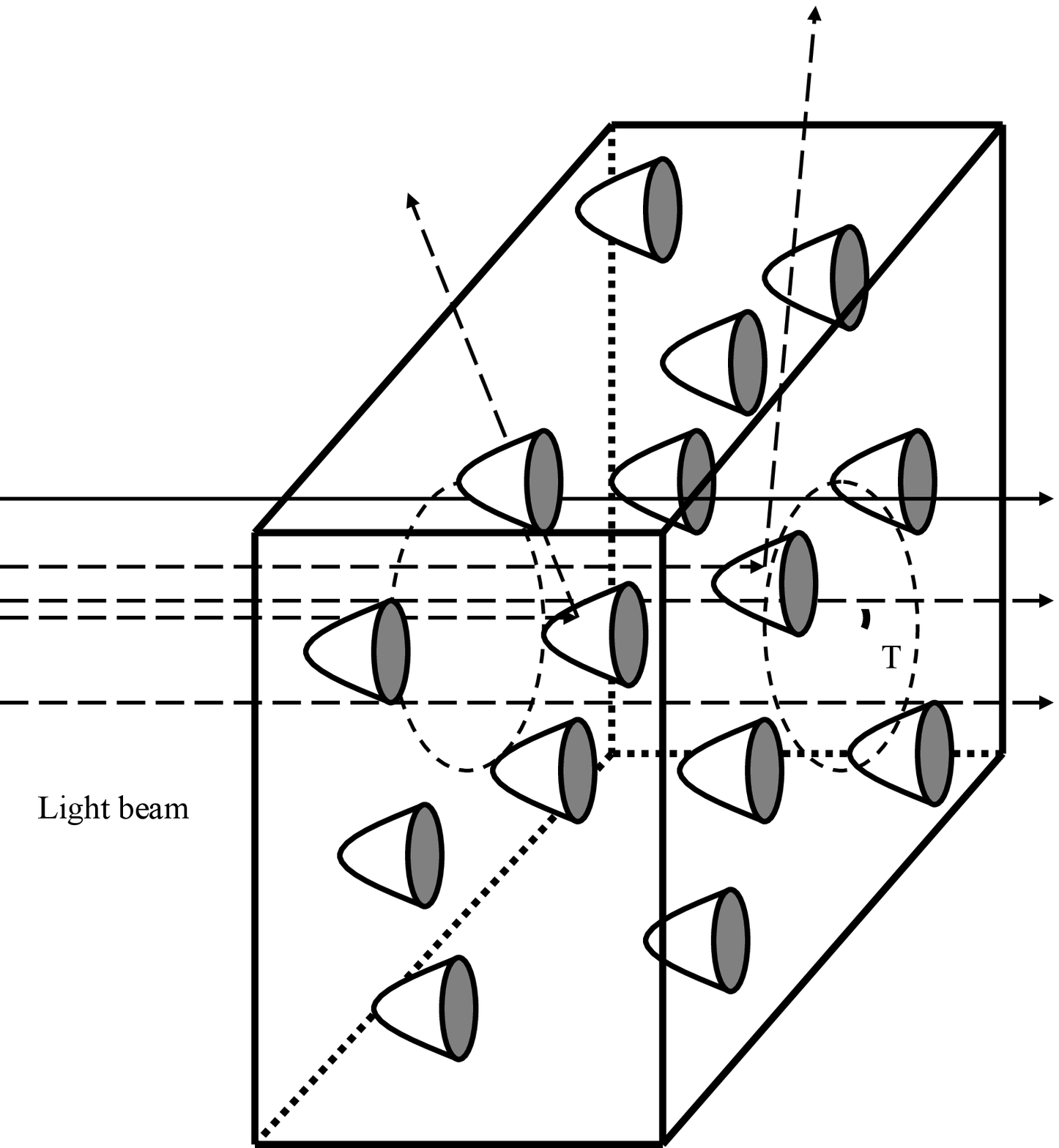,width=0.9\linewidth}
\caption{A collimated beam is shot against a thin target of reflecting solids, each one of total cross-section $\sigma_T$, fixed at some points of the space. Most of the beam will be undeflected but part of it will be scattered by the solids.}
\label{fig1}
\end{figure}

Suppose we have some {\it perfectly reflecting} solids, held fixed
at some points of space, and imagine to shoot at them a well
collimated light beam  whose wavelength is short with respect to
the dimensions of these solids, and which is diffused by the
individual targets: see figure \ref{fig1}.

A first question we can ask ourselves is: what fraction of the beam
hits the targets? Or, more precisely worded, what is the ratio of
scattered energy flux to incoming energy flux? It is evident that
the answer to the above question depends:

a) on the number of targets per unity volume, that is on the
density $n$ of the targets,

b) on the length $h$ of the layer of targets,

c) and on the section $\sigma_T$ that  each one of them shows to the beam.

This last quantity is what is called the total cross-section for the beam-target interaction.

Let us suppose, for simplicity, that:

i) the beam has section $S$;

ii)  the individual targets are sufficiently spaced from each other
and $h$ is small, so that $nh\sigma_T \ll 1$ and the probability
that one target covers another is negligible (thin target
hypothesis).

In this case the total area shown to the beam by the target
contained in the volume of base $S$ and height $h$ is
$nSh\sigma_T$; and the fraction $P$ of the beam hitting the targets
is given by its ratio to the beam section $S$:
\begin{equation}
P={\frac{nSh\sigma_T}{S}}=nh\sigma_T\label{one}
\end{equation}
Provided the target is thin, in the sense of hypothesis ii),
relation (\ref{one}) may be regarded as a general definition of the
total cross-section $\sigma_T$.

In the following we shall always assume hypothesis ii) to be
verified. For simplicity we shall moreover only consider individual
targets having axial symmetry and assume the beam to be shot at
them along the direction of their axis.

\section{The concept of differential cross section}\label{tre}

As the individual targets are perfectly reflecting, the incoming
beam is {\it deflected} by the interaction with the target. A
second question we can therefore ask ourselves is: what fraction of
the incoming beam is scattered by more than a given angle?

To answer this question, we can start considering a beam ray
hitting one of the individual targets, at a distance $b$ from the
axis of the body ($b$ is called impact parameter). The ray will be
deflected according to the reflection laws: the incident ray, the
perpendicular to the reflecting surface at the reflection point and
the reflected ray all lie in the same plane and the incidence angle
is equal to the reflection angle. Let us call $\phi$ the scattering
angle, that is the angle of deviation from the incident beam
direction caused by the reflection; see figure \ref{fig2}.

One can see that if the target is convex (the second derivative of
the generating curve is strictly positive), then the smaller the
impact parameter $b$ the larger is the scattering angle $\phi$.
That means that all the rays within a disc of area $\pi b^2$, with
center on the target axis, and perpendicular to their direction,
will suffer a scattering angle greater than $\phi$; see again
figure \ref{fig2}. We can now answer our question by saying that
the fraction of the incoming beam which is scattered through an
angle greater than $\phi$ is:
\begin{equation}
P_{>\phi}=nh\pi b^2
\end{equation}

In other  words and keeping in mind equation (\ref{one}), it can be
said that the cross-section for scattering through an angle greater
than $\phi$ is:
\begin{equation}
\sigma_{>\phi}= \pi b^2\label{two}
\end{equation}

We observe that equation (\ref{two}) is not ``fundamental" in that
it relates the cross-section to the impact parameter $b$ which, in
a scattering experiment, where the positions of the individual
targets are not known, is not a measurable quantity. Nonetheless
this equation will be of great importance for the next
considerations.

We now further refine the question to: what fraction of the beam is
scattered through an angle between $\phi$ and $\phi+d\phi$?

The rays scattered through an angle between $\phi$ and $\phi+d\phi$
are given by those deflected through an angle greater than $\phi$
minus those deflected more than $\phi+d\phi$. These are the rays
that hit the fixed body on a cross surface of area
$|d\sigma|=\sigma_{>\phi}-\sigma_{>(\phi+d\phi)}$; the required
fraction is then:
\begin{equation}
P(\phi)= nh|d\sigma|
\end{equation}

After the scattering, these rays are contained into a solid angle
of amplitude $d\Omega$ given by the ratio between the area of the
spherical zone $Z$, of figure \ref{fig3}, and $r^2$, that is:
\begin{equation}
d\Omega= {\frac{2\pi(r\sin\phi)rd\phi}{r^2}}=2\pi\sin\phi d\phi\label{three}
\end{equation}

and therefore the the beam fraction scattered around the angle
$\phi$, per unit solid angle, is:
\begin{equation}
P(\phi)= nh{\frac{|d\sigma|}{d\Omega}}\label{four}
\end{equation}

Equation (\ref{four}) is of general interest and it i's valid for all
scattering experiments in the thin target hypothesis. Introducing
the incoming flux per unit surface $dP/dS$ and the scattered flux
per unit sold angle $dP/d\Omega$, it can be recast in the more
symmetric form:
\begin{equation}
{\frac{dP}{d\Omega}} =nhS{\frac{|d\sigma|}{d\Omega}}{\frac{dP}{dS}}.
\end{equation}

The fundamental quantity $|d\sigma|/d\Omega$ is called the
differential cross-section. To better understand  its physical
meaning we can extend our description of the experiment to include
the detection process. Consider a beam shot against a fixed target
and an ideal detector of effective section $A$, located at an angle
$\phi$ at a distance $R$ from the target and perpendicular to the
scattered beam direction. In this way it detects all the radiation
in the solid angle of amplitude $\Omega \sim A/R^2$. If $\Omega$ is
sufficiently small, $P(\phi)$ can be considered to be constant over
the surface $A$, and the ratio between the detected and incident
beam flux is given by $P(\phi)$ multiplied by $\Omega$ which is:
\begin{equation}
P_{\Omega}(\phi)= nh{\frac{|d\sigma|}{d\Omega}}{\frac{A}{R^2}}\label{five}
\end{equation}

Equation (\ref{five}) shows a clear way of calculating the
differential cross-section from given experimental measures and a
comparison between (\ref{four}) and (\ref{five}) helps students
enlighten the conceptual meaning of this useful quantity.

\section{Detailed calculation}\label{qua}

\begin{figure}[htbp]
\centering\epsfig{file=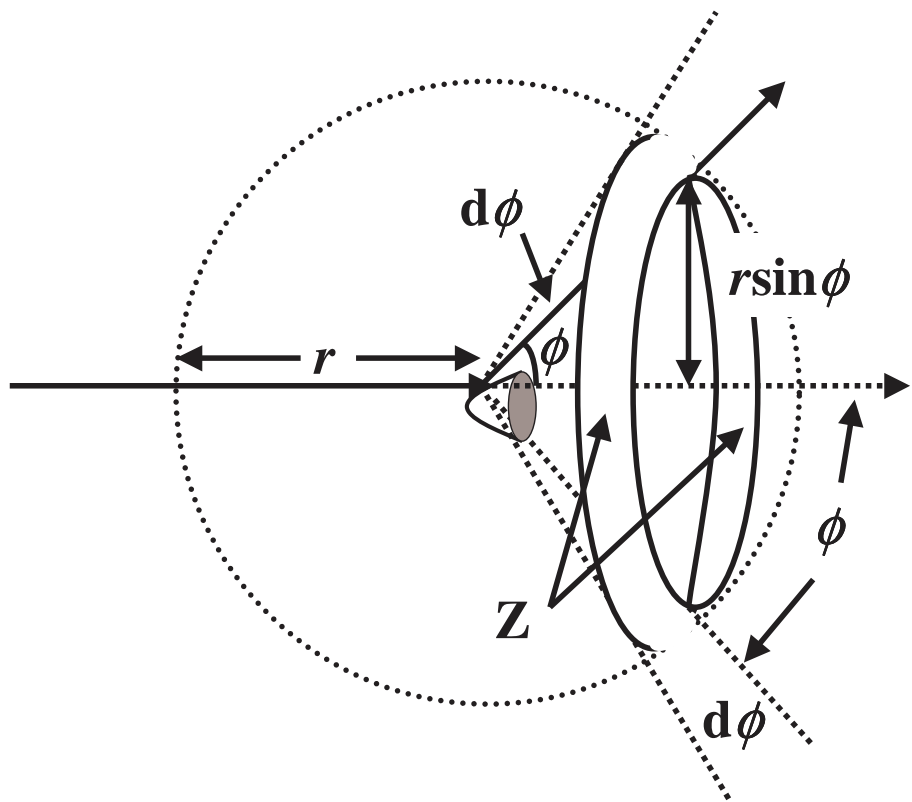,width=0.9\linewidth}
\caption{A sharp-edged solid is given by the complete rotation around the y axis of the increasing convex function  $y = f(x)$ with $x$ between $0$ and $a$. A beam ray, directed along the $y$ axis, is reflected and scattered by the solid through the angle $\phi$: $b$ is the impact parameter, $t$ the tangent and $p$ the perpendicular to the solid at the collision point, $c$ and $d$ are respectively the incident and the reflected ray.}
\label{fig2}
\end{figure}

\begin{figure}[htbp]
\centering\epsfig{file=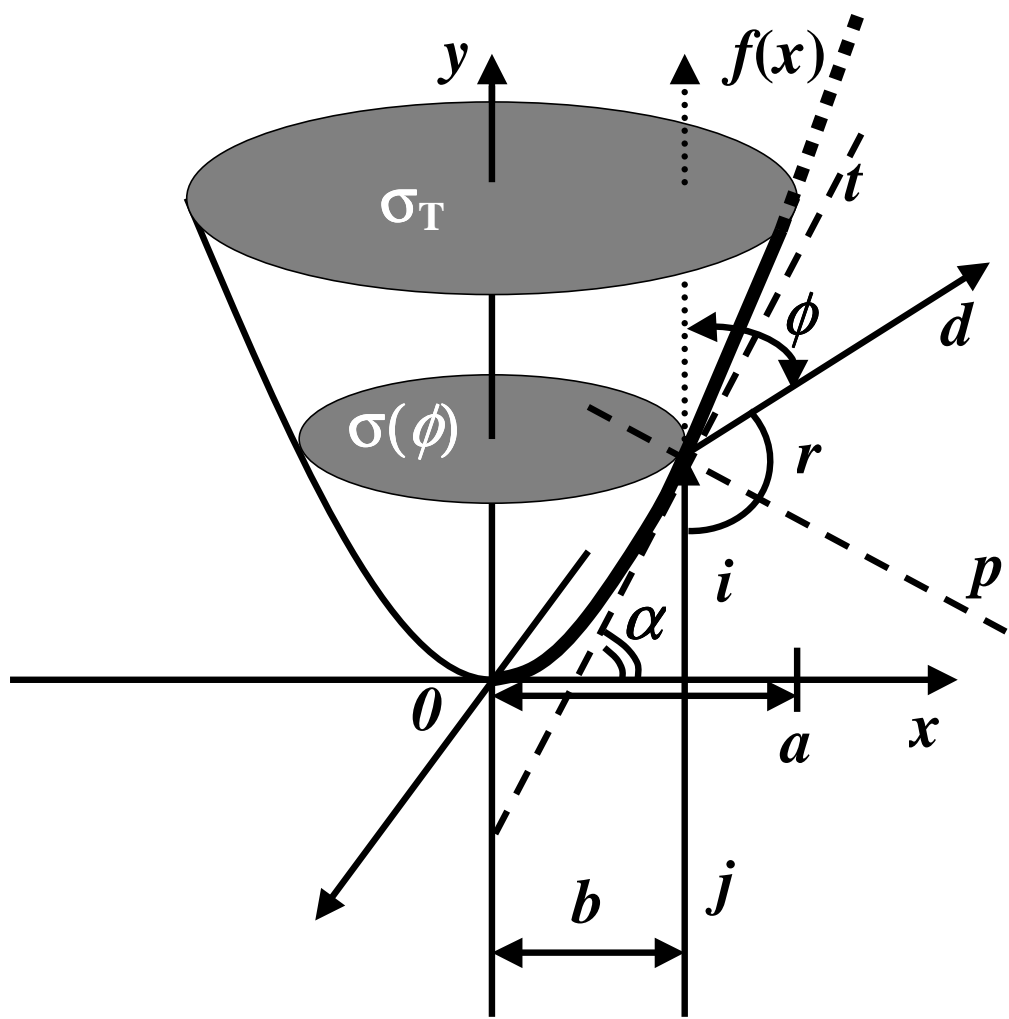,width=0.9\linewidth}
\caption{The beam rays scattered between $\phi$  and $\phi +d\phi$  are contained into a solid angle of amplitude given by the ratio between the area $2\pi r^2\sin\phi d\phi$, of the spherical zone $Z$, and $r^2$.}
\label{fig3}
\end{figure}

Let's consider the rigid solid produced by the complete rotation
around the $y$ axis of the increasing convex function $y = f(x)$
with $x$ between $0$ and $a$; figure \ref{fig2}. Our aim is to
calculate the total and differential cross-sections for the
reflection of a beam, incident on this fixed solid, in the
direction of the $y$ axis. The calculation of the total
cross-section
\begin{equation}
\sigma_T=\pi a^2 \label{six}
\end{equation}
is straightforward. For what concerns the differential
cross-section we make reference to figure \ref{fig2}: since the
incident ray, the perpendicular to the surface and the reflection
ray lie all on the same plane, the problem can be considered in the
$x-y$ plane. Let $b$ be the impact parameter, $t$ the tangent to
the curve in the collision point, $p$ the perpendicular, $j$ the
line of incidence and $d$ the line of reflection. Now $j$ is
orthogonal to the $x$ axis and $p$ is orthogonal to $t$, therefore
the angle $\alpha$, between $t$ and the $x$ axis, is equal to the
angle ${\hat{i}}$ between $j$ and $p$, which is the incidence angle.

Keeping an eye on figure \ref{fig2}, it follows that the connection between the impact parameter $b$ and
the deflection angle $\phi$ is:
\begin{equation}
g(b)\equiv\left.{\frac{df}{dx}}\right|_{x=b}
=\tan{\alpha}=\tan{\hat{i}}=\tan\left({\frac{\pi-\phi}
2}\right)=\cot{\frac{\phi}{2}} \label{seven}
\end{equation}
where $\left.{\frac{df}{dx}}\right|_{x=b}$ is the derivative of $f$ at point $b$.

We are now able to obtain the differential cross-section. The principal steps are the following:

{\bf I.} Solve equation (\ref{seven}) with respect to $b$, that is:
express $b$ as a function of $\phi$, with the obvious geometric
limitation $0\leq b \leq a$ due to the finite size of the target.
In formulae, defining the inverse function $g^{-1}$ through the
equation $g^{-1}(g(x))=x$:
\begin{eqnarray}
\left\{
\begin{array}{l}
b(\phi)=g^{-1}\left(\cot{\frac{\phi}{2}}\right)\\
\pi-2\arctan(g(a))\leq \phi \leq \pi-2\arctan(g(0))%
\end{array}%
\right.\label{eight}
\end{eqnarray}

{\bf II.} Calculate $\sigma(\phi)=\pi b^2$.

{\bf III.} Differentiate $\sigma(\phi)$ to obtain $|d\sigma(\phi)|$.

{\bf IV.} Divide the found expression by $d\Omega$ given by eq.(\ref{three}).

\section{Three examples}\label{cin}

We present here, as examples of the procedure above, three cases we
think significant, together with sample discussions of some key
points of the theory; teachers can choose among them, make up their
own comments, or invent other examples more suited to their
individual aims.

\begin{figure}[htbp]
\centering\epsfig{file=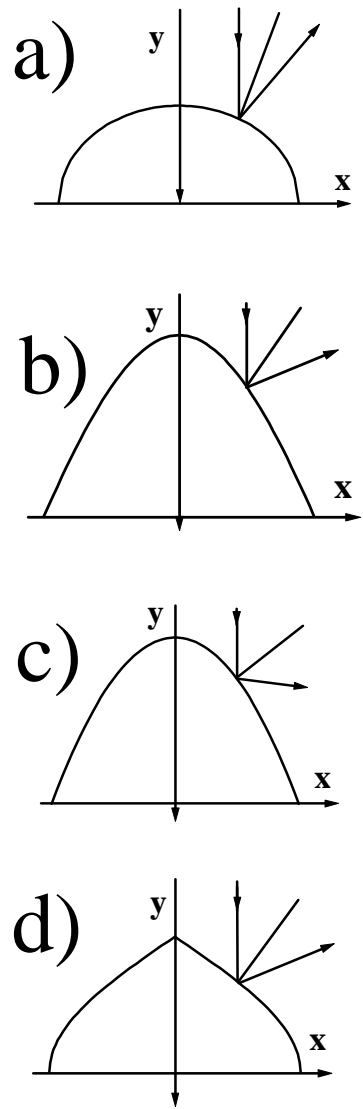,width=0.4\linewidth}
\caption{The curves generating the reflecting solids of our examples: a) an ellipse; b) an hyperbola; c) a parabola; d) an inverse sine.}
\label{fig4}
\end{figure}

{\bf 1) ellipsoids, paraboloids and spheres:} let's consider the
ellipsoid obtained by the rotation, around the $y$ axis, of the
function (shown in Fig. \ref{fig4} a)
\begin{equation}
f(x)=-c\sqrt{1-\left({\frac{x}{a}} \right)^2};\hspace{.8in}0\leq x \leq a,\label{nineb}
\end{equation}
with $a$ and $c$ the two semi-axes of the generating ellipse (this example is discussed in detail in
\cite{cor}).

From step I, equation (\ref{eight}), we get
\begin{eqnarray}
b^2=a^2{{\cot^2\left({{\phi}\over 2}\right)}\over{{{c^2}\over {a^2}}+\cot^2\left({{\phi}\over 2}\right)}},
\end{eqnarray}
and then (steps II., III., and IV.):
\begin{eqnarray}
\left\{
\begin{array}{l}
{\frac{|d\sigma(\phi)|}{d\Omega}} = {{a^2}\over
4}\left({\frac{ac}{c^2\sin^2\left({\frac{\phi}{2}}\right)+a^2\cos^2\left({\frac{\phi}{2}}\right)}}\right)^2.\label{nine} \\ 0\leq \phi \leq \pi%
\end{array}%
\right.
\end{eqnarray}

From the obvious relations: $\int{d\Omega}=4\pi$ and
$\int{{{d\sigma}\over{d\Omega}} d\Omega}=\sigma_T$, where both
integrals are on the whole solid angle (or on ``all directions"),
and from equation (\ref{nine}), one gets immediately the trivial
result $\sigma_T  = \pi a^2$. This can be a useful check for most
students.

A simple change of signs in eq. (\ref{nineb}) gives us the equation
of a hyperboloid (shown in Fig. \ref{fig4} b):
\begin{equation}
f(x)=c\sqrt{1+\left({x \over {\tilde a}} \right)^2};\hspace{.8in}0\leq x \leq a,\label{ninec}
\end{equation}
which results in the cross-section
\begin{eqnarray}
\left\{
\begin{array}{l}
{\frac{|d\sigma(\phi)|}{d\Omega}} = {\frac{{\tilde a}^2}{4}}\left({\frac{{\tilde a}c}{c^2\sin^2\left({\frac{\phi}{2}}\right)-{\tilde a}^2\cos^2\left({\frac{\phi}{2}}\right)}}\right)^2. \\
\pi-2\arctan\left({\frac{ca} {\tilde a}}{\frac{1}{\sqrt{{\tilde a}^2+a^2}}}\right)\leq \phi \leq \pi%
\end{array}%
\right.
\end{eqnarray}
where, as the curve (\ref{ninec}) is not bound in $x$, we had to
distinguish between the target size parameter $a$ (introduced as to
avoid an infinite total cross-section) and the curve parameter
${\tilde a}$.

Changing the ratio $c/a$ for the ellipsoid, or $c/{\tilde a}$ for
the hyperboloid, changes the angular distribution of the scattered
rays. In particular, for $c=a$, eq. (\ref{nine}) reduces the well
known expression of the differential cross-section of a rigid
sphere \cite{lan} $|d\sigma(\phi)|/d\Omega =a^2/4$, which depends
only on the sphere radius $a$, which is a constant, but is
independent of $\phi$. This means that after the collision with a
sphere the rays are isotropically scattered, that is: they are
deflected to every angle with equal probability. This result
depends on the particular interaction here considered.

In a scattering experiment we do not usually know the {\it exact}
nature and composition of the target, and the aim of the experiment
is to deduct the missing information, be it the potential for a
target composed of force centers, or the generating curve $f(x)$ in
the cases here discussed. Direct reconstruction of the unknown
scattering potential, or of the generating curve, is an arduous
task, on which whole books have been written \cite{sab}; on the
other hand, partial information on the nature of the target can
help reduce the range of possible interactions. This either makes a
``trial and error" approach, such as was used in the interpretation
of early scattering experiments, more feasible, or even allows
univocal direct solution of the problem \cite{varie}.

The simplest example one can give -that of isotropic scattering-
already allows for the highlight of some necessary subtleties: if
in a scattering experiment the detector gives the same reading at
all angles, independently from the physical properties of the
incident beam, such as, for example, the energy (such a dependence
would suggest a target composed of force centers \cite{lan}), or
the electric charge carried by the beam, from which it could a
priori depend, we have strong indications that the interaction
between the beam and the target is a reflection off the surface of
the individual target, and that the target is ``made of" rigid
spheres, impenetrable by the beam.

It seems a simple and straightforward procedure, but we must note
that both the angular dependence, {\it and also its range of
validity} are important to determine the scattering surface: a
differential cross-section which over a given range of angles has
the constant value $a^2/4$ can be caused by targets generated by
any curve having a derivative of the form \cite{int}
\begin{equation}
g(x)=\sqrt{{\gamma^2+x^2}\over{\beta^2-x^2}},\hspace{.1in}0\leq x\leq \beta,\hspace{.1in}\gamma^2+\beta^2=a^2,
\end{equation}
but it is zero outside the range $0\leq \phi \leq
\pi-2\arctan(|\gamma / \beta|)$, while the differential
cross-section for spheres is constant over the whole range $0\leq
\phi \leq \pi$.

More examples of reconstruction of hard reflecting surfaces of
revolution from observed cross sections can be found in Ref.
\cite{brun}.

{\bf 2) Paraboloids:} let us consider the solid obtained by the
rotation of the curve \cite{evans} (shown in Fig. \ref{fig4} c)
\begin{equation}
f(x)={{x^2}\over c};\hspace{.8in}0\leq x \leq a.
\end{equation}
Reminding equation (\ref{six}), the total cross section is again
$\sigma_T =\pi a^2$. For the differential cross-section, step I.
(equation (\ref{eight})) instead gives us:
\begin{eqnarray}
b ={c \over 2}\cot\left({{\phi}\over 2}\right)
\end{eqnarray}
The following steps (steps II.to IV.) then give us:
\begin{eqnarray}
\left\{
\begin{array}{l}
\frac{{|d\sigma|}{d\Omega}} = {\frac{\pi c^2} {4}}{\frac{\cos\left({\frac{\phi}{2}}\right)d\phi}{\sin^3\left({\frac{\phi}{2}}\right)}}
{\frac{1}{2\pi\sin\phi d\phi}}={\frac{c^2} {16}}
cosec^4\left({\frac{\phi} {2}}\right).\label{eleven}\\
\pi-2\arctan\left({2{\frac{a} {c}}}\right)\leq \phi \leq \pi%
\end{array}%
\right.
\end{eqnarray}

Note that, contrary to the previous case, where the ratio $c/a$
influences the angular distribution of the scattered beam, here $c$
is just a scale factor: the angular distribution is independent
from it.

This result shows some similarities with the Rutherford
cross-section \cite{ruth,will}:
\begin{equation}
{\frac{|d\sigma_R|}{d\Omega}} = {\frac{K^2} {16}}
cosec^4\left({\frac{\phi}{2}}\right);\hspace{.9in}K={\frac{Ze\rho_q}{4\pi\epsilon_0
\rho_{K_0}}},
\end{equation}
where $Ze$  is the target charge, $\rho_q$ and $\rho_{K_0}$ are the
charge and kinetic energy densities of the incident beam, and
$\epsilon_0$ is the vacuum permittivity \cite{closest}. Clearly,
the same angular dependence appears; moreover, both of the  cross
sections are not valid for small angles, but for very different
reasons: in the present case because of the discontinuity of $f(x)$
in $x=a$, and in the Rutherford case because when the impact
parameter is large the shielding effect of the atoms' electronic
cloud is no more negligible. How can we then distinguish the two
cases?  First of all, eq. (\ref{eleven}) is only valid when the
incidence direction is that of the symmetry axis of the individual
targets; a rotation of the target with respect to the incident beam
would result in a different result. More important, the
differential cross-section given by eq. (\ref{eleven}) shows no
dependence from the kinetic energy carried by the incident beam
(the variable in the Rutherford experiment which can be most easily
changed): we would therefore have to imagine different scattering
surfaces for beams having different kinetic energies: a highly
unlikely circumstance.

{\bf 3) the $\arcsin$ as the symmetric scatterer to the Paraboloid:}
let's finally consider the surface obtained by the rotation, around
the $y$ axis, of the function (shown in Fig. \ref{fig4} d)
\begin{equation}
f(x)={\widetilde c}\arcsin {x \over a};\hspace{.8in}0\leq x \leq a.
\end{equation}

From step I, equation (\ref{eight}), we get
\begin{eqnarray}
b^2=a^2-{\widetilde c}^{\hspace{.02in}2} \tan^2\left({{\phi}\over 2}\right)
\end{eqnarray}
then (steps II.- IV.):
\begin{eqnarray}
\left\{
\begin{array}{l}
{\frac{|d\sigma|} {d\Omega}} = {\frac{{\widetilde c}^{\hspace{.02in}2}}
{4}}{\frac{1}{\cos^4\left({\frac{\phi}{2}}\right)}}
={{{\widetilde c}^{\hspace{.02in}2}}\over 4}
sec^4\left({{\phi}\over 2}\right)\\ 0\leq \phi \leq
\pi-2\arctan\left({{\widetilde c} \over a}\right),%
\end{array}%
\right.
\end{eqnarray}
which is a mirror image of our previous example eq. (\ref{eleven}),
obtained by the substitutions $\phi \rightarrow \pi
- \phi$ and $c \rightarrow 2{\widetilde c}$. The transformation
between the two ranges in $\phi$ follows the same simple rule, as
for $\alpha \in (0,\pi/2)$,
\begin{equation}
{{\pi}\over 2} - \arctan \alpha = \arctan \left({1 \over {\alpha}}\right).
\end{equation}

\section{Discussion and Conclusions}\label{sei}

The concept of cross-section in the case of reflection off the
surface of a solid has been introduced on the basis only of simple
statistical and geometric considerations. This choice has been made
because of didactic reasons: with this approach, the intuition is
helped by the ``material" existence of the cross-sections both
total and differential (which are real parts of the rigid surface
of the body).

In physically more significant cases, as e. g. the Rutherford
scattering itself, the interaction is between the incoming beam and
the force field of the target. It is true that in these cases the
total cross-section can be geometrically conceived as the
``effective surface" presented by the target force field to the
incoming beam and the differential cross section as the ``part" of
that ``surface" scattering the beam in a given direction, but this
only in an abstract sense, which usually is not easy for the
students to grasp: for example, even in the simple geometrical case
of a light beam randomly scattered by a rough surface, the
differential cross section, even if mathematically defined, cannot
be associated with any geometrically definite part of the
scattering surface.

Besides this difficulty, the general approach, in which the
scattering is introduced from the beginning in its abstract
meaning, often allows students only to discuss existing
experimental data, thus missing important relations and theoretical
considerations, an understanding of which can be gained by the
direct calculation of several simple cross-sections.

In conclusion, this paper gives a simple formula to calculate with
little effort many different cross-sections, three of which are
explicitly given (in the case of a rigid ellipsoid, in that of a
rigid paraboloid, and in that of the solid generated by the
rotation of an inverse sine curve). The formula itself is of
limited utility because of the very simple interaction taken into
account but, at the introductory level, it can help students
understand the link between experiments and theoretical
explanations.

An extended case study, about the students' understanding of the
cross-section concept, following the ideas outlined in the present
paper, is under way and will be presented in a forthcoming paper.

\ack

We are very grateful to Professors Ludovico Lanz, Lucio Rossi and
Guido Vegni for very helpful discussions and their comments on the
manuscript. 
%The present work was conducted within the frame of the
%Italian inter-university project ``F21 - Percorsi di formazione in
%fisica per il $21^o$ secolo".

\end{document}